\documentclass[twocolumn,prb,superscriptaddress,longbibliography]{revtex4-2}

\usepackage{amsmath,amsfonts,amssymb}
\usepackage{graphicx,color}
\usepackage{hyperref}
\usepackage{epstopdf}
\usepackage{float}
\usepackage{multirow}
\usepackage{hhline}
\usepackage{ulem}
\usepackage{mathtools}
\usepackage[caption=false]{subfig}
\usepackage{soul}

\usepackage{graphicx}
\usepackage{dcolumn}
\usepackage{bm}

\usepackage[utf8]{inputenc} 

\usepackage{multirow}
\usepackage[warn]{mathtext}
\usepackage{amssymb}

\usepackage{textcomp} 
\usepackage{indentfirst} 
\usepackage{amsmath} 
\usepackage{graphicx}
\DeclareGraphicsExtensions{.pdf,.png,.jpg}
\usepackage{pgfplots}
\pgfplotsset{compat=1.13}

\usepackage{hyperref}
\hypersetup{
    colorlinks=true,
    linkcolor=blue,
    filecolor=black,      
    urlcolor=blue,
    citecolor= violet
}

\usepackage{algpseudocode}

\usepackage{adjustbox}
\usepackage{tabularx}
\usepackage[braket, qm]{qcircuit}

\usepackage{mathtools}

\usepackage{xcolor}
\definecolor{C0}{HTML}{4C72B0}
\definecolor{C1}{HTML}{DD8452}
\definecolor{C2}{HTML}{55A868}
\definecolor{C3}{HTML}{C44E52}

\usepackage{ulem}



\newcolumntype{Y}{>{\centering\arraybackslash}X}

\makeatletter
\def\@fnsymbol#1{\ensuremath{\ifcase#1\or \dagger\or \ddagger\or
   \mathsection\or \mathparagraph\or \|\or **\or \dagger\dagger
   \or \ddagger\ddagger \else\@ctrerr\fi}}
\makeatother

\footnotetext{These authors contributed equally to this work.}

\begin{document}

\title{Native CCZ Gate with Fluxonium Qubits and a Microwave-Driven Coupler}

\author{Grigoriy S. Mazhorin$^*$}
\email[Corresponding author: ]{mazhorin.gs@phystech.edu}
\affiliation{National University of Science and Technology ``MISIS'', 119049 Moscow, Russia}
\affiliation{ International Center for Quantum Optics and Quantum Technologies, Assn, 127051 Moscow, Russia}

\author{Tatyana A. Chudakova$^*$}
\affiliation{National University of Science and Technology ``MISIS'', 119049 Moscow, Russia}
\affiliation{ International Center for Quantum Optics and Quantum Technologies, Assn, 127051 Moscow, Russia}
\affiliation{Moscow Center for Advanced Studies, 123592 Moscow, Russia}

\author{Alena S. Kazmina}
\affiliation{National University of Science and Technology ``MISIS'', 119049 Moscow, Russia}
\affiliation{ International Center for Quantum Optics and Quantum Technologies, Assn, 127051 Moscow, Russia}

\author{Nikolai~G.~Berezkin}
\affiliation{National University of Science and Technology ``MISIS'', 119049 Moscow, Russia}
\affiliation{ International Center for Quantum Optics and Quantum Technologies, Assn, 127051 Moscow, Russia}
\affiliation{Moscow Center for Advanced Studies, 123592 Moscow, Russia}

\author{Arina~V.~Zotova}
\affiliation{National University of Science and Technology ``MISIS'', 119049 Moscow, Russia}
\affiliation{ International Center for Quantum Optics and Quantum Technologies, Assn, 127051 Moscow, Russia}
\affiliation{Moscow Center for Advanced Studies, 123592 Moscow, Russia}

\author{Artyom~M.~Polyanskiy}
\affiliation{National University of Science and Technology ``MISIS'', 119049 Moscow, Russia}
\affiliation{ International Center for Quantum Optics and Quantum Technologies, Assn, 127051 Moscow, Russia}
\affiliation{Moscow Center for Advanced Studies, 123592 Moscow, Russia}

\author{Nikolay~N.~Abramov}
\affiliation{National University of Science and Technology ``MISIS'', 119049 Moscow, Russia}

\author{Mikhail~A.~Tarkhov}
\affiliation{Institute of Nanotechnology of Microelectronics, Russian Academy of Sciences, Moscow, 119991 Russia}

\author{Alexander~M.~Mumlyakov}
\affiliation{Institute of Nanotechnology of Microelectronics, Russian Academy of Sciences, Moscow, 119991 Russia}

\author{Igor~V.~Trofimov}
\affiliation{Institute of Nanotechnology of Microelectronics, Russian Academy of Sciences, Moscow, 119991 Russia}

\author{Elizaveta~A.~Krivko}
\affiliation{Institute of Nanotechnology of Microelectronics, Russian Academy of Sciences, Moscow, 119991 Russia}

\author{Nikita~Yu.~Rudenko}
\affiliation{National University of Science and Technology ``MISIS'', 119049 Moscow, Russia}

\author{Maxim~V.~Chichkov}
\affiliation{National University of Science and Technology ``MISIS'', 119049 Moscow, Russia}

\author{Vladimir~I.~Chichkov}
\affiliation{National University of Science and Technology ``MISIS'', 119049 Moscow, Russia}

\author{Ilya A. Simakov}
\affiliation{National University of Science and Technology ``MISIS'', 119049 Moscow, Russia}
\affiliation{ International Center for Quantum Optics and Quantum Technologies, Assn, 127051 Moscow, Russia}

\date{\today}

\begin{abstract}

Native multi-qubit gates could reduce the overhead associated with decompositions into single- and two-qubit operations, but whether they can simultaneously provide high fidelity, simple control and robustness against parasitic interactions in scalable architectures remains unclear.
Here we experimentally realize a 65-ns native controlled-controlled-phase operation, locally equivalent to the Toffoli gate, with a fidelity of 99.39(5)\% in a three-qubit processor unit based on fluxonium qubits coupled via a microwave-driven transmon coupler.
The implemented operation would require CZ fidelities of approximately 99.94\% if realized through a conventional decomposition.
The gate is implemented with a single control pulse, 
that relies on a simple calibration procedure yielding coherence-limited performance. 
This processor unit naturally extends to scalable two-dimensional layouts with low parasitic interactions.
Altogether, these results establish native multi-qubit gates as a viable hardware-efficient primitive for scalable superconducting quantum processors.
\end{abstract}

\maketitle

Quantum processors based on native single- and two-qubit gates have achieved remarkable progress, establishing a scalable platform for quantum computation~\cite{Arute_2019, google2023suppressing, Abanin2025}. The predominance of two-qubit entangling gates is largely driven by hardware considerations rather than algorithmic requirements: these operations offer simple and robust calibration, high-fidelity implementation, and enable large-scale processors with suppressed parasitic interactions~\cite{acharya2024quantumerrorcorrectionsurface, kim2023evidence, gao2025establishing}. Many quantum algorithms, in contrast, rely on multi-qubit primitives, whose decompositions into two-qubit gates require multiple entangling steps, longer execution times and accumulated gate errors~\cite{PhysRevLett.121.010501, mukhopadhyay2023synthesizing, sun2024quantum}. This overhead raises a fundamental question: can a processor be designed in a way to facilitate multi-qubit entangling operations as a native computational building block, while retaining the simplicity, robustness and scalability of conventional two-qubit approaches?

The pursuit of such native multi-qubit primitives has recently attracted growing interest across several quantum-computing platforms, including trapped ions~\cite{monz2009realization, nikolaeva2025scalable}, neutral atoms~\cite{levine2019parallel, evered2023high}, and spin-based systems~\cite{hendrickx2021four, takeda2022quantum}. In superconducting circuits, a natural route is to extend architectures based on pairwise interactions by more complex control schemes to realize effective three-qubit gates within the same hardware framework~\cite{lvb9-pfr3, nguyen2024programmable, kim2022high}. 
Such approaches preserve scalability and suppression of parasitic interactions and also support continuous families of three-qubit gates~\cite{gu2021fast,warren2023extensive}. 
However, this flexibility comes at the cost of elaborate calibration involving multiple simultaneous drives, leading to increased coherent errors that constrain performance even as coherence times improve, eliminating the advantage over decomposed implementations.
An alternative strategy relies on direct many-body interactions, enabling entangling operations on three~\cite{PhysRevApplied.14.014072} or more qubits~\cite{song2017continuous,PhysRevLett.128.190502} in a single step.  While conceptually simpler, current realizations exhibit strong parasitic interactions limiting scalability and complicating single-qubit control.
Thus, no existing superconducting architecture with native multi-qubit gates simultaneously achieves the simplicity, high fidelity, and scalability of conventional two-qubit-based approaches.

For a native multi-qubit operation to provide a computational advantage, it must be both algorithmically important and expensive to be synthesized from two-qubit gates.
The three-qubit Toffoli operation is a key primitive for classical reversible logic~\cite{toffoli1980reversible} and central to quantum algorithms, including Grover search~\cite{PhysRevLett.79.325, figgatt2017complete}, Shor factoring~\cite{shor1999polynomial, haner2017factoring}, the Harrow-Hassidim-Lloyd algorithm~\cite{PhysRevLett.103.150502}, and quantum chemistry simulations~\cite{PRXQuantum.5.040339}, error-correction protocols, enabling measurement-free syndrome processing~\cite{PRXQuantum.5.010333,butt2025measurement}. Yet its decomposition requires eight two-qubit gates for nearest-neighbour connectivity~\cite{smith2023leap}, resulting in substantial error accumulation. 
An efficient native realization of the Toffoli gate or its equivalent would therefore provide a relevant step towards establishing multi-qubit gates as practical primitives in scalable superconducting architectures.

\begin{figure*}
    \centering
    \includegraphics{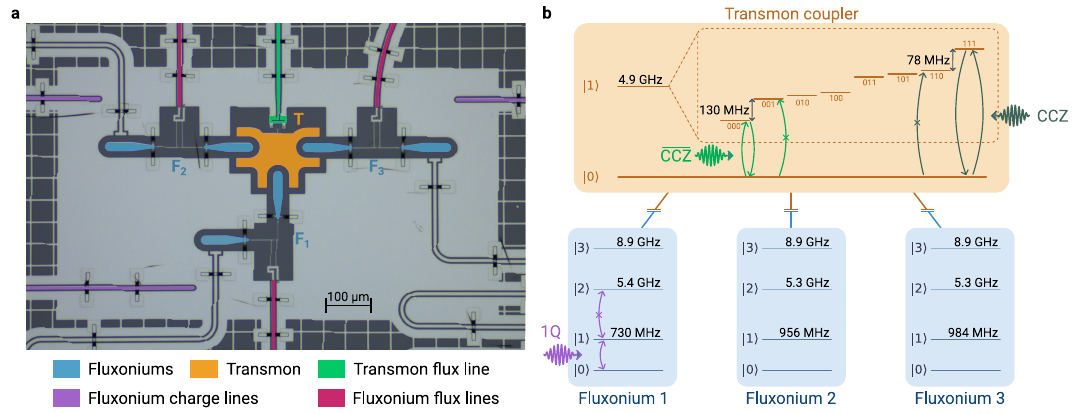}
    \caption{\textbf{Device and three-qubit gate concept.}
\textbf{a}, False-colour micrograph of the quantum processor comprising three fluxonium qubits (blue) capacitively connected to a transmon coupler (orange). Individual flux-bias lines (pink), charge-drive lines (violet), and readout resonators are coupled to each fluxonium qubit for flux tuning, single-qubit control, and dispersive state measurement, respectively. The native three-qubit gate is driven by a single microwave pulse applied to the coupler flux line (green).
\textbf{b},  
System energy levels.
The blue insets show the bare energy levels of the individual fluxoniums, while the orange inset shows the transmon levels conditioned on the computational fluxonium states. Arrows indicate the microwave-driven transitions used for single-qubit (violet) and three-qubit operations (bright and dark green for $\overline{\mathrm{CCZ}}$ and $\mathrm{CCZ}$, respectively). Crossed arrows denote the nearest unwanted transitions that limit spectral selectivity: the $\ket{1}-\ket{2}$ fluxonium transition for single-qubit gates and the closest transmon $\ket{0}-\ket{1}$ transition associated with an untarget computational state for the three-qubit gates. The corresponding frequency separations are determined by the fluxonium anharmonicity and the state-dependent transmon frequency shifts (130~MHz for $\overline{\mathrm{CCZ}}$ and 78~MHz for CCZ).}
    \label{fig:design}
\end{figure*}

Here we experimentally demonstrate an elementary three-qubit processor unit featuring a native three-qubit CCZ gate locally equivalent to a Toffoli operation. The device is based on fluxonium~\cite{manucharyan2009fluxonium} qubits strongly capacitively connected through a transmon~\cite{PhysRevA.76.042319} coupler, whose transition frequency therefore depends on the joint computational state of neighbouring qubits, enabling spectral selectivity for the entangling operation~\cite{proposal_CCZ, simakov2023coupler, FTF_MIT}.
The implemented CCZ gate is activated by a single microwave $2\pi$ pulse and admits a simple calibration procedure based on two standard independent measurements. 
We measure a fidelity of 99.39(5)\% at a duration of 65~ns. 
Achieving the same fidelity with a conventional decomposition on processors with nearest-neighbour connectivity would require CZ gate fidelities exceeding 99.9\%.
The gate mechanism also enables independent and simultaneous control of phase accumulation on the $\ket{000}$ and $\ket{111}$ computational states using a multiplexed microwave drive.
We further show that the processor unit naturally extends to scalable two-dimensional layouts while preserving the mechanisms used to suppress parasitic interactions in linear fluxonium--transmon architectures~\cite{kugut2025interaction, zhan2026scalable}. Together, these results demonstrate that native multi-qubit operations can be realized without sacrificing the simplicity, robustness, and scalability traditionally associated with superconducting processors based on two-qubit gates, paving the way towards their practical use as computational primitives.

\section*{Three-qubit processor unit}

\begin{table}[t]
\centering
\caption{\textbf{Device parameters.}
The table summarizes the measured characteristics of the three-qubit processor unit, including qubit coherence properties, static ZZ couplings, and state-dependent coupler transition frequencies.
F$_1$--F$_3$ denote fluxonium qubits and T denotes the transmon coupler.}
\label{tab:exp_params}
\begin{tabular}{lcccc}
\hline
\hline
 & \textbf{F1} & \textbf{F2} & \textbf{F3} & \textbf{T} \\
\hline
$\omega_{01}/2\pi$ (GHz)        & 0.730  & 0.956  & 0.984  & 4.901 \\
$T_1$ ($\mu$s)                  & 92    & 90    & 75    & 6 \\
$T_2^\text{echo}$ ($\mu$s)             &  17   & 10    &  12    & 6 \\
\hline
\hline
\multicolumn{5}{c}{\textbf{Static fluxonium--fluxonium interactions}} \\
\hline
$ZZ_{12}/2\pi$ (kHz)            & \multicolumn{4}{c}{9.8 ± 1.6} \\
$ZZ_{13}/2\pi$ (kHz)            & \multicolumn{4}{c}{9.2 ± 3.8} \\
$ZZ_{23}/2\pi$ (kHz)            & \multicolumn{4}{c}{8.7 ± 1.7} \\
\hline
\end{tabular}
\begin{tabular}{cccccccc}
\hline
\multicolumn{8}{c}{\textbf{State-dependent coupler frequency}} \\
\multicolumn{8}{c}{\textbf{(detunings from 4.901 GHz, in MHz)}} \\
\hline
$\ket{000}$ & $\ket{001}$ & $\ket{010}$ & $\ket{011}$ & $\ket{100}$ & $\ket{101}$ & $\ket{110}$ & $\ket{111}$ \\
\hline
$0$& $130$&$132$&$238$&$228$&$310$&$310$&$388$\\
\hline
\hline
\end{tabular}
\end{table}

Our processor unit (Fig.~\ref{fig:design}a) consists of three computational fluxonium qubits F$_1$--F$_3$ capacitively coupled via a shared transmon coupler T. Key device parameters are summarized in Table~\ref{tab:exp_params}.
The architecture combines strong fluxonium--transmon coupling with suppressed parasitic interactions between computational qubits.
Conditional Ramsey oscillations yield a longitudinal ZZ interaction below 10~kHz.
This behaviour originates from the large fluxonium anharmonicity, which keeps the low-energy qubit subspace spectrally isolated while enabling coupling to a high-frequency transmon via non-computational fluxonium transitions~\cite{FTF_MIT}.
As a result, the coupler transition frequency acquires a pronounced dependence on the joint state of the qubits.
This leads to frequency-resolved coupler transitions for all computational basis states (Fig.~\ref{fig:design}b), providing the selectivity that enables the entangling gate.

At a phenomenological level, this behaviour can be described by an effective longitudinal interaction between qubits and coupler of the form:
\begin{equation}
\frac{\hat{H}_{\mathrm{eff}}}{\hbar}
=
\sum_{i=1}^3\frac{\omega_i}{2}\hat{\sigma}_z^{(i)}
+ 
\left(
\frac{\omega_T}{2}
+
\sum_{i=1}^3\frac{\chi_i}{4}\hat{\sigma}_z^{(i)}
\right)
\hat{\sigma}_z^{(T)},
\end{equation}
where $\omega_i$ are the fluxonium transition frequencies, $\omega_T$ is the coupler frequency, and $\chi_i$ are state-dependent frequency shifts.
This minimal model captures the mechanism underlying the gate and provides an intuitive description of the state-dependent coupler response, with a more detailed treatment given in the Supplementary Information.
In our device, the coupling strength is comparable to the detuning between the fluxonium 1--2 transition and the coupler 0--1 transition, leading to non-additive shifts (Table~\ref{tab:exp_params}).

State-selective excitation of the coupler directly implements a CCZ gate. 
A near-resonant $2\pi$ pulse applied at the transmon frequency associated with the $\ket{111}$ computational state selectively induces a full Rabi oscillation on the corresponding dressed transition (Fig.~\ref{fig:design}b). 
The driven evolution returns the coupler to its ground state and imparts a phase of $\pi$ to the $\ket{111}$ state, thereby realizing a native three-qubit CCZ gate. 
Physically, this operation is equivalent to a single-qubit rotation of the transmon coupler and therefore inherits the conceptual simplicity of single-qubit control, requiring only a single microwave pulse and admitting straightforward calibration.

An entirely analogous protocol enables conditional phase accumulation on the $\ket{000}$ computational state, implementing three-qubit operation, equivalent to a CCZ gate up to single-qubit $X$ rotations, which we denote throughout the paper as $\overline{\mathrm{CCZ}}$.
In the present device, coupler transition associated with the $\ket{000}$ exhibits better spectral isolation, and the following experimental results therefore focus on the $\overline{\mathrm{CCZ}}$ implementation.

\section*{High-fidelity native CCZ gate}

\begin{figure}
    \centering
    \includegraphics[width=\linewidth]{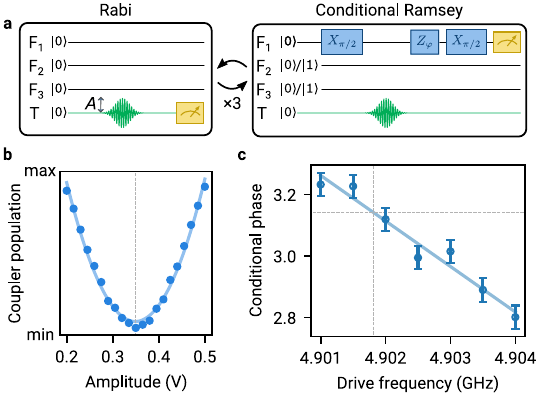}
    \caption{ \textbf{Calibration of the native three-qubit gate. a,} Pulse sequences used for the Rabi-like amplitude calibration and the Ramsey-type conditional-phase calibration. The pulse sequences are shown for the $\overline{\mathrm{CCZ}}$ gate; for the $\mathrm{CCZ}$ gate, the initial fluxonium states are inverted.
\textbf{b,} Measured coupler excited-state population as a function of the drive amplitude. The dashed line indicates the minimum population corresponding to a complete $2\pi$ rotation of the coupler.
\textbf{c,} Measured conditional phase as a function of the drive frequency. The vertical dashed line marks the calibrated pulse frequency, while the horizontal dashed line indicates the target phase of $\pi$.}
    \label{fig:calib}
\end{figure}

Consistent with the single-qubit-like nature of the gate, the implemented CCZ operation requires only two independent calibration steps for a fixed gate duration and pulse shape: Rabi-like and Ramsey-type measurements of the pulse amplitude and drive frequency, respectively. In the former, for target computational state the drive amplitude is swept to identify the value corresponding to a $2\pi$ rotation by minimizing the residual coupler population after the pulse. In the latter, the drive frequency is swept while monitoring the conditional phase accumulated on one of the qubits. The Ramsey-type procedure is performed for several initial qubit states to determine the conditional phase. The calibration cycle is repeated three times. The pulse sequences and representative results are shown in Fig.~\ref{fig:calib}. Further details of the conditional-phase extraction and calibration-error analysis are provided in the Supplementary Information.

\begin{figure*}
    \centering
    \includegraphics[width=\linewidth]{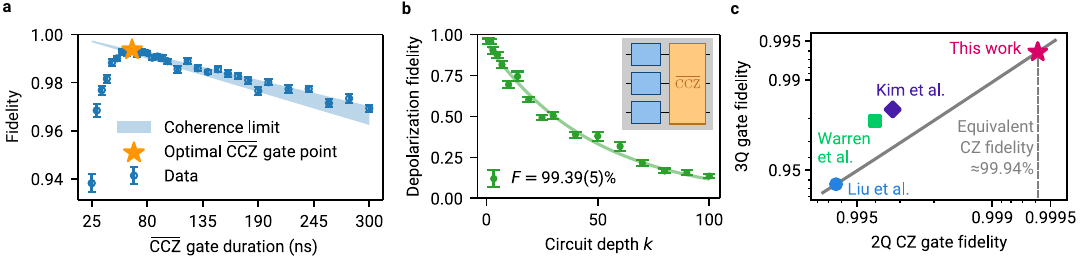}
    \caption{\textbf{Benchmarking of the native $\overline{\mathbf{CCZ}}$ operation. a,} $\overline{\mathrm{CCZ}}$ gate fidelity extracted from XEB as a function of gate duration. The blue shaded region indicates the coherence limit estimated from independent coherence-time measurements. The orange star marks the best observed fidelity of $99.39(5)\%$ at a gate duration of $65$ ns. 
    \textbf{b,} XEB of the optimal $\overline{\mathrm{CCZ}}$ gate together with the corresponding XEB circuit. Blue gates denote randomly sampled single-qubit Clifford operations.
    \textbf{c,} Performance comparison of the native three-qubit gate demonstrated in this work with direct state-of-the-art realizations of other three-qubit gates: CCZ, Liu et al.~\cite{lvb9-pfr3}; iToffoli, Kim et al.~\cite{sun2024quantum}; and CCZS, Warren et al.~\cite{warren2023extensive}. The gray line shows the numerically estimated effective CCZ fidelity obtained from decompositions based on two-qubit gates of varying fidelity. The vertical dashed line indicates the equivalent CZ gate fidelity corresponding to the measured $\overline{\mathrm{CCZ}}$ fidelity.}
    \label{fig:results}
\end{figure*}

The gate duration is not intrinsically constrained by the underlying mechanism and can be varied continuously, enabling optimization of the trade-off between coherent and incoherent errors.
Fig.~\ref{fig:results}a shows the measured gate performance as a function of pulse length obtained using cross-entropy benchmarking (XEB) with a single-qubit references~\cite{simakov2026decay}, achieving a maximum fidelity of $99.39(5)\%$ at a gate duration of 65~ns.
The observed dependence exhibits two features of the pulse-duration behaviour of microwave-driven single-qubit gates: a rapid drop at short pulses and an approximately linear decrease at longer durations.
For fast drives, the performance degrades due to unwanted off-resonant excitation of nearby transmon transitions arising from the finite spectral width of the pulse.
In this respect, the spectral isolation of the target transition plays the same role as the qubit anharmonicity in single-qubit gates, setting the level of coherent errors.
At long durations, the gate fidelity approaches the decoherence limit, in agreement with independent estimates of qubit energy relaxation and dephasing.
This trend indicates that the gate performance is primarily limited by qubit coherence times rather than by intrinsic constraints of the gate mechanism.
Numerical simulations support this interpretation, predicting errors below $10^{-4}$ for gate durations around 100~ns in the absence of decoherence~\cite{proposal_CCZ}.

\begin{figure}
    \centering
    \includegraphics[width=\linewidth]{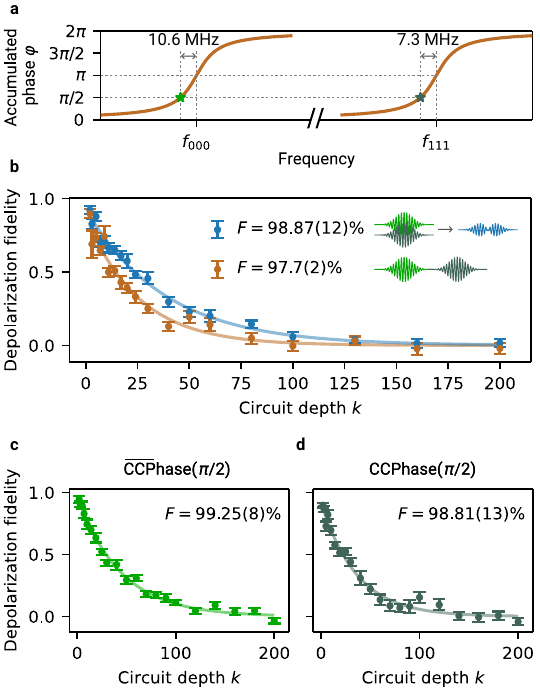}
    \caption{
    \textbf{Conditional phase control.}
    \textbf{a,} Conditional phase as a function of the drive frequency. $f_{000}$ and $f_{111}$ denote the transmon $\ket{0}\!-\!\ket{1}$ transition frequencies conditioned on the $\ket{000}$ and $\ket{111}$ computational states of the fluxoniums, respectively. The blue and green stars indicate the drive frequencies used to implement the $\pi/2$ conditional phase on the $\ket{111}$ and $\ket{000}$ states, respectively.
    \textbf{b,} Comparison between the multiplexed gate performance (blue) and its sequential implementation (brown). 
    \textbf{c,} XEB of the 80-ns $\mathrm{\overline{CCP}hase}(\pi/2)$ and $\mathrm{CCPhase}(\pi/2)$ gates. 
    }
    \label{fig:family}
\end{figure}

Native implementations are expected to be most beneficial for gates that are expensive to synthesize from one- and two-qubit operations. For linearly connected qubits, a Toffoli-equivalent $\overline{\mathrm{CCZ}}$ circuit requires eight CZ gates~\cite{nielsen2010quantum}, making it a compelling target for hardware-level realization.
To quantify the advantage of the native implementation, we estimate the effective fidelity of the corresponding decomposition using the superconducting-inspired SI1000 circuit-level noise model while varying the underlying CZ fidelity (see Supplementary Information). The resulting dependence is shown in Fig.~\ref{fig:results}c.
The measured $\overline{\mathrm{CCZ}}$ fidelity of $99.39(5)\%$ corresponds to that expected from a decomposition employing two-qubit gates with fidelities of 99.94\%, at the level of the best reported superconducting two-qubit gates~\cite{Manucharyan_24_days, marxer2026above}.
This result arises from reaching the coherence-limited regime at a gate duration comparable to a two-qubit operation, avoiding multi-step circuit execution.
The demonstrated three-qubit gate performance therefore becomes competitive with decomposed implementations realized on qubits with coherence times around an order of magnitude longer.
Further improvements in coherence should directly translate into higher three-qubit gate fidelity.

\section*{Multiplexed control}

The gate mechanism naturally enables simultaneous and  independent control of the phases accumulated on the $\ket{000}$ and $\ket{111}$ computational states using a multiplexed coupler drive. 
Fig.~\ref{fig:family}a shows how each pulse component selectively addresses the corresponding dressed coupler transition while the others remain well detuned, allowing continuous phase control of the target states without introducing additional coherent errors.

As an illustration, we realize a three-qubit gate imparting phases of $\pi/2$ to both the $\ket{000}$ and $\ket{111}$ states.  
We compare the multiplexed and the corresponding sequential implementation using XEB. Fig.~\ref{fig:family}b shows the depolarization fidelity as a function of the number of applied three-qubit gates for two sequential pulses (brown) and a multiplexed pulse (blue). The extracted gate fidelities are 97.7(2)\% and 98.87(12)\%, respectively.
The multiplexed pulse is constructed as the direct sum of two individually calibrated control-phase drives. Due to the limited output power of the arbitrary waveform generator, the comparison is performed for 80~ns pulses.
The higher fidelity of the multiplexed approach results from its shorter execution time, demonstrating the benefit of simultaneous phase control.

To verify the absence of additional coherent errors from multiplexed control, we separately benchmark the constituent $\mathrm{\overline{CCP}hase}(\pi/2)$ and $\mathrm{CCPhase}(-\pi/2)$ gates (Fig.~\ref{fig:family}c). The difference in their fidelities is attributed to unequal spectral isolation of the coupler transitions associated with the $\ket{000}$ and $\ket{111}$ states (78 MHz for the CCZ gate and 130 MHz for the $\overline{\mathrm{CCZ}}$ gate implementation as shown in Fig.~\ref{fig:design}b). Accordingly, the fidelity of the simultaneous implementation is limited by the lower of the two individual fidelities. 

These results further demonstrate the robustness of the gate mechanism against coherent errors, as simultaneous driving with individually calibrated control parameters does not introduce observable additional errors.

\section*{Discussion}

The advantages of the native multi-qubit operation can be fully realized only if the underlying gate mechanism supports scaling to large devices exhibiting low parasitic interactions. In fluxonium--transmon circuits with strong capacitive coupling, the use of differential fluxoniums together with alternating transmon frequencies~\cite{kugut2025interaction} is sufficient to suppress unwanted interactions in one-dimensional architectures. 
This approach has recently been employed in a 22-qubit linear fluxonium processor~\cite{zhan2026scalable}. 
In two-dimensional lattices multiple transmons are inevitably connected to the same fluxonium island, introducing direct capacitive coupling between them. 
This issue may be addressed by auxiliary elements~\cite{kugut2025interaction} or by complex couplers such as the double-transmon~\cite{chan2026system}. 
The three-qubit coupling element allows each fluxonium island to remain connected to at most one transmon even in two-dimensional layouts, naturally extending the interaction-suppression mechanism of linear processors without additional circuit complexity.

Three-qubit-gate architectures further preserve another feature of its microwave-activated CZ based counterparts~\cite{kugut2025interaction}: the ability to perform simultaneous entangling gates acting on sharing qubits. 
Such parallelism arises from driving adjacent couplers targeting transitions associated with different computational states, directly analogous to the multiplexed control demonstrated here.
The only difference is that both drive tones are applied to a single coupling element rather than to neighbouring ones.
We experimentally observe independent phase accumulation on the $\ket{000}$ and $\ket{111}$ states without observable fidelity degradation, indicating that simultaneous excitation of distinct coupler transitions does not introduce additional interaction conflicts and supporting the feasibility of fully parallel native three-qubit operations in larger-scale processors.

The results presented here demonstrate that native three-qubit gates can combine simple control, low coherent errors, and clear scaling strategy.
Using a single microwave pulse, we realize a CCZ operation with a fidelity of 99.39(5)\%, required equivalent CZ gate fidelities of about 99.94\% for decomposition.
Implemented multiplexed control further highlights the flexibility of the underlying gate mechanism.
Our work shows that scalable superconducting processors can be designed around native many-body interactions without sacrificing hardware efficiency. 
Such architectures could reduce the execution-time overhead associated with decompositions of multi-qubit operations by providing a more direct hardware implementation of algorithmically relevant quantum primitives.

\section*{Acknowledgements} 

We thank Alexey Ustinov, Ilya Besedin and Ilya Moskalenko for helpful discussions and critical comments on the manuscript. 
The work was supported by the Ministry of Science and Higher Education of the Russian Federation in the framework of the Program of Strategic Academic Leadership “Priority 2030” (MISIS Strategic Technology Project Quantum Internet).

\section*{Author Contributions} 
G.S.M. and I.A.S. conceived and supervised the project.
G.S.M. designed the device.
M.A.T., A.M.M., I.V.T., E.A.K., N.Y.R., M.V.C., and V.I.C. fabricated the device.
T.A.C., A.S.K., A.M.P., and N.N.A. prepared the experimental setup. 
T.A.C. conducted the experiment with help from G.S.M., A.S.K. and I.A.S. 
G.S.M., T.A.C., A.S.K., N.G.B., A.V.Z, and I.A.S. analyzed the data.
G.S.M., I.A.S. prepared the manuscript with contribution of all authors.

\section*{Competing interests}
The authors declare no competing interests.

\section*{Supplementary Information}

\appendix

\section{Circuit concept}
\begin{figure}
    \centering
    \includegraphics[width=\linewidth]{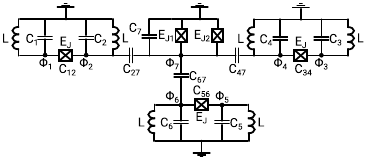}
    \caption{\textbf{Equivalent circuit of the processor.}
}
    \label{fig:scheme}
\end{figure}

Here we derive the Hamiltonian of the three-qubit processor whose equivalent circuit is shown in Fig.~\ref{fig:scheme}.
The processor consists of three two-mode fluxonium qubits~\cite{moskalenko2021tunable} coupled via a frequency-tunable transmon coupler. Each two-mode fluxonium comprises two superconducting islands shunted by superinductances ($L$) implemented as Josephson-junction arrays and connected by a small Josephson junction ($E_J$). Two Josephson junctions ($E_{J1}$ and $E_{J2}$) form an asymmetric SQUID of the transmon, providing in-situ frequency tunability.

The use of two-mode fluxonium qubits is motivated by several practical considerations. For a fixed charging energy, the two-mode design provides larger electrode capacitances, reducing the relative impact of parasitic capacitances while leaving sufficient space for coupling to other circuit elements, including the transmon coupler, readout resonators, and control lines. Unlike floating designs~\cite{FTF_MIT}, it also allows galvanic coupling to the flux-bias line, enabling stable operation at half-flux bias without requiring large bias currents.

The tunable coupler frequency can be used to compensate for fabrication-induced frequency variations and to avoid resonant coupling to spurious two-level systems. In the present experiment, however, such corrections were not required, and the coupler was operated at its lower sweet spot throughout the measurements.

The capacitance matrix of the complete circuit, $\mathbf C$, is extracted from electromagnetic simulations. The diagonal elements represent the sum of the self-capacitance and all mutual capacitances connected to the corresponding node, whereas the off-diagonal elements equal the negative mutual capacitances between the two nodes.
The dominant capacitances  are
$C_{1}=C_{2}=C_{3}=C_{4}=24.5$~fF,
$C_{12}=C_{34}=1.2$~fF,
$C_{5}=C_{6}=24.3$~fF,
$C_{56}=1.4$~fF,
$C_{7}=69$~fF,
$C_{27}=C_{47}=C_{67}=6.4$~fF.

The circuit Lagrangian reads
\begin{equation}
\mathcal L
=
\frac{1}{2}
\dot{\boldsymbol{\Phi}}^{T}
\mathbf C
\dot{\boldsymbol{\Phi}}
-
\Pi .
\end{equation}

Here $\boldsymbol{\Phi} =
(\Phi_1,\Phi_2,\Phi_3,\Phi_4,\Phi_5,\Phi_6,\Phi_7)^T$,
where $\Phi_i$ denotes the flux associated with the $i$-th circuit node and the potential energy is
\begin{equation}
\Pi = \sum_{i=1}^3 \Pi_{F_i} + \Pi_T .
\end{equation}

It is convenient to express the Lagrangian in terms of superconducting phases $\varphi=\frac{2\pi}{\Phi_0}\Phi$, where $\Phi_0$ is magnetic flux quantum.

Using new variables, for fluxoniums
\begin{equation}
\Pi_{F_i}
=
- E_{J_F}\cos(\varphi_{2i-1}-\varphi_{2i}-\varphi_{F_i}^{\mathrm{ext}})
+ \frac{E_{L}}{2}
(\varphi_{2i-1}^2+\varphi_{2i}^2),
\end{equation}
and for the tunable transmon
\begin{equation}
\Pi_T
=
- E_{J1}\cos\varphi_7
-
E_{J2}\cos(\varphi_7-\varphi_T^{\mathrm{ext}}).
\end{equation}
where $\Phi_i^{\mathrm{ext}} = \frac{\Phi_0}{2\pi}\varphi_i^{\mathrm{ext}}$ denotes the external magnetic flux threading the corresponding fluxonium loop or transmon SQUID loop.

Since the Josephson potential depends only on the phase difference across the junctions, we introduce auxiliary variables

\begin{equation}
\theta_i = \varphi_{2i-1} + \varphi_{2i},
\qquad
\eta_i = \varphi_{2i-1} - \varphi_{2i}.
\end{equation}

Thus

\begin{equation}
\boldsymbol{\psi}
=
(\theta_1,\eta_1,\theta_2,\eta_2,\theta_3,\eta_3,\varphi_7)^T
=
\mathbf S \boldsymbol{\varphi},
\end{equation}
with transformation matrix

\begin{equation}
\mathbf S =
\begin{pmatrix}
1 & 1 & 0 & 0 & 0 & 0 & 0 \\
1 & -1 & 0 & 0 & 0 & 0 & 0 \\
0 & 0 & 1 & 1 & 0 & 0 & 0 \\
0 & 0 & 1 & -1 & 0 & 0 & 0 \\
0 & 0 & 0 & 0 & 1 & 1 & 0 \\
0 & 0 & 0 & 0 & 1 & -1 & 0 \\
0 & 0 & 0 & 0 & 0 & 0 & 1
\end{pmatrix}.
\end{equation}

The capacitance matrix transforms as

\begin{equation}
\mathbf C' = (\mathbf S^{-1})^{T} \mathbf C \mathbf S^{-1}.
\end{equation}

In the new variables the fluxonium potential becomes

\begin{equation}
\Pi_{F_i}
=
- E_{J_F}\cos(\eta_i-\varphi_{F_i}^{\mathrm{ext}})
+ \frac{1}{4}E_L(\theta_i^2+\eta_i^2).
\end{equation}

Introducing Cooper-pair number operators

\begin{equation}
\mathbf n
=\frac{1}{2e}
\frac{\partial \mathcal L}{\partial \dot{\boldsymbol{\psi}}},
\qquad
\mathbf E_C = \frac{e^2}{2} \mathbf C'^{-1},
\end{equation}
and applying Legendre transformation, the Hamiltonian takes the form:

\begin{equation}
H
=
4\mathbf n^{T}
\mathbf E_C
\mathbf n
+
\sum_{i=1}^3 \Pi_{F_i}
+
\Pi_T .
\label{eq:full_Ham}
\end{equation}

The diagonal elements of $\mathbf E_C$ define the effective charging energies of the fluxonium qubits, the harmonic modes, and the transmon, while the off-diagonal elements describe the corresponding capacitive couplings.
The additional harmonic mode associated with the two-mode fluxonium architecture~\cite{moskalenko2021tunable} is engineered to remain far detuned from both the qubit and coupler transitions. Consequently, it neither participates in the gate dynamics nor introduces significant parasitic interactions, and is therefore omitted from the effective model used in the remainder of this work.

\section{Device parameters}
The effective Hamiltonian derived from Eq.~\eqref{eq:full_Ham}, which captures the relevant dynamics of the quantum processor, reads
\begin{equation}
H
=
\sum_{i=1}^3 H_{F_i}
+
H_T
+
\frac{1}{2}\sum_{i,j} \mathrm{g}_{ij} \, n_i n_j,
\label{eq:rest_Ham}
\end{equation}
where $H_{F_i}$ describes the fluxonium mode $\eta_i$ of the $i$-th qubit and $H_T$ the transmon. In the interaction term, the indices $i$ and $j$ run over all modes, including the three fluxonium modes $F_i$ and the transmon $T$.

The fluxonium Hamiltonian is
\begin{equation}
H_{F_i}
=
4E_{C,i} n_{F_i}^2
-
E_{J_F}\cos(\eta_i-\varphi_{F_i}^{\mathrm{ext}})
+
\frac{1}{2}E_L \eta_i^2,
\end{equation}
while the transmon Hamiltonian reads
\begin{equation}
H_T
=
4E_{C,T} n_T^2
-
E_{J1}\cos\varphi_7
-
E_{J2}\cos(\varphi_7-\varphi_T^{\mathrm{ext}}).
\end{equation}

To determine the parameters of the effective Hamiltonian, we perform two complementary sets of measurements. The fluxonium parameters are extracted from two-tone spectroscopy as a function of the applied external flux. The conditional transmon $0$--$1$ transition frequencies are measured for all computational states of the fluxoniums at both zero transmon flux and half-flux bias, while keeping the fluxoniums at half-flux bias. The transition frequencies are determined by sweeping the drive frequency using a pulse whose amplitude and duration are calibrated to produce a resonant $\pi$ rotation.

The measured frequencies are fitted using the Hamiltonian model with free parameters corresponding to qubit energies and coupling strengths. 
In the numerical simulation, each subsystem is first diagonalized independently, and the full Hamiltonian is then constructed in the truncated basis, retaining five lowest levels for each fluxonium mode and three lowest levels for the transmon.
The extracted parameters are summarized in Table~\ref{tab:spec_params}.

\begin{table}[h]
\centering
\caption{\textbf{Device parameters.}
The table summarizes theextracted parameters of the three-qubit processor unit, including qubit charging, inductive and Josephson energies and coupling coefficients.
F$_1$--F$_3$ denote fluxonium qubits and T denotes the transmon coupler.}
\label{tab:spec_params}
\begin{tabular}{lcccc}
\hline
\hline
 & $\textbf{F}_\textbf{1}$ & $\textbf{F}_\textbf{2}$ & $\textbf{F}_\textbf{3}$ & \textbf{T} \\
\hline
$E_L/h$ (GHz)        & 1.06  & 1.08  & 1.07  & - \\
$E_J/h$ (GHz)                  & 5.1    & 4.45    & 4.38    & 16.46 and 4.59 \\
$E_C/h$ (GHz)             &  11.82   & 11.76    &  11.83    & 2.31 \\
\hline
\hline
\multicolumn{5}{c}{\textbf{Coupling strengths $\textbf{g}_{\textbf{\textit{{ij}}}}$ (MHZ) }} \\
\hline
 & $F_1$ & $F_2$ & $F_3$ & T \\
\hline
$F_1$        & -  & 129  & 132  & 547 \\
$F_2$                  & 129    & -    & 132    & 546 \\
$F_3$            &  132   & 132    &  -    & 559 \\
T            &  547   & 546    &  559    & - \\
\hline
\hline
\end{tabular}
\end{table}

\section{Fabrication}
The three-qubit quantum processor is fabricated in three stages. First, the surface of high-resistivity silicon substrate is cleaned using wet chemical cleaning methods in strong acid and alkali solutions (SC1, Piranha), followed by removal of the natural silicon oxide layer using BOE (buffered oxide etchant). A 100 nm thick aluminum film is deposited using electron-beam evaporation. The topological pattern of the high-Q resonators and readout lines are defined using maskless laser lithography (with AZ1505 photoresist) and subsequent transferred by plasma-chemical etching of the functional aluminum layer~\cite{chudakova2024effect}. After the topological pattern transfer, the photoresist is removed using wet chemical removal methods in solvents.

In the second stage, Al/AlOx/Al Josephson junctions are formed using shadow evaporation on a bilayer mask and lift-off lithography using Dolan bridge technology. The substrate with formed high-Q resonator structures is deposited with bilayer mask composed of a stack of MMA EL6/AR-P 6200.09 electron resists. The Josephson junction structures are exposed and developed in two stages, followed by treatment of the bilayer mask in oxygen plasma. Aluminum electrodes are deposited on the formed bilayer mask at two different angles with intermediate oxidation. The Josephson junctions are finally formed using lift-off lithography by soaking the substrate in an NMP (N-methyl-2-pyrrolidone) solution for three hours at a temperature of $80^\circ\mathrm{C}$.

At the third stage, the peripheral components of multi-qubit quantum circuits were formed: bandages and ground potential equalization elements. The bandage formation technology is carried out according to the technological implementation given in the work~\cite{10.1063/1.4993577}, the technology for forming the ground potential equalization elements is carried out according to the technological implementation given in the work~\cite{10.1063/1.4863745}.

\section{Experimental setup}
\label{sec:experimental_setup}
\begin{figure}[t]
    \centering
    \includegraphics[width=\linewidth]{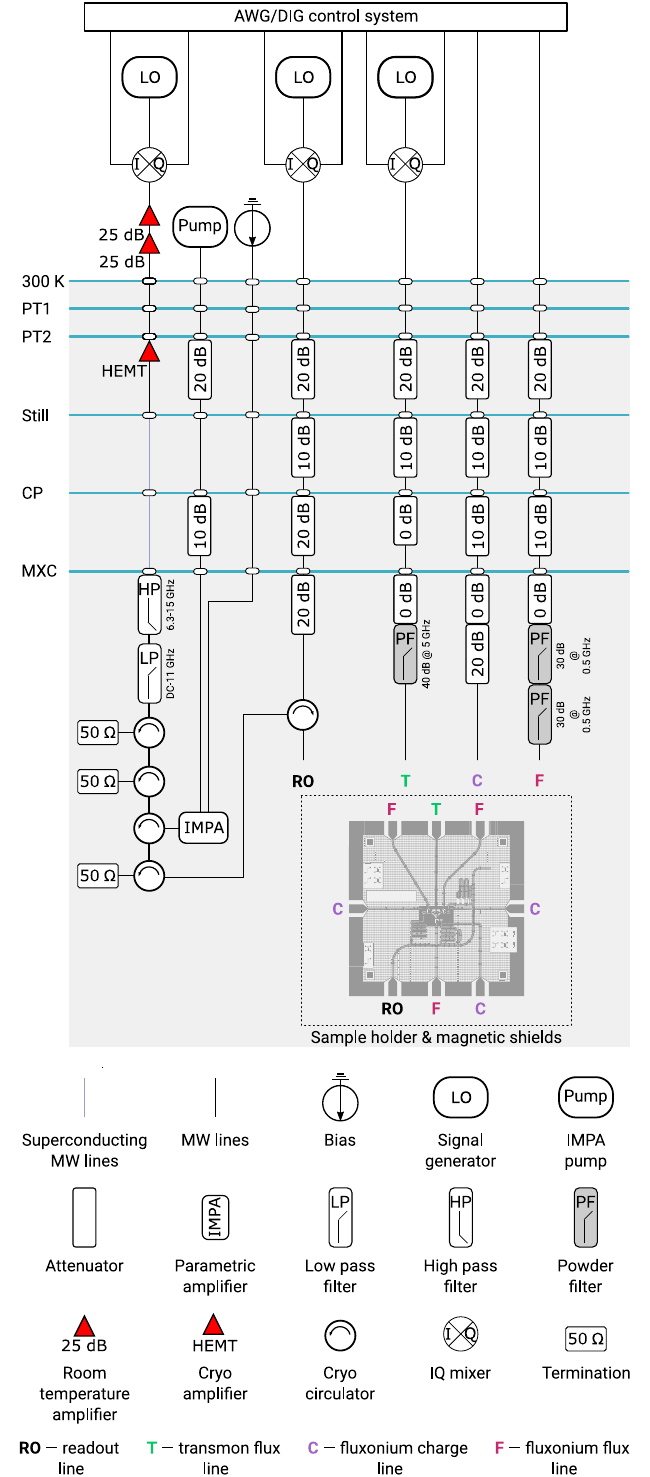}
    \caption{\textbf{Experimental setup.} 
    }
    \label{fig:experimental_setup}
\end{figure}


All measurements are performed in a dilution refrigerator with a base temperature of approximately 20~mK. A schematic of the complete experimental setup is shown in Fig.~\ref{fig:experimental_setup}. At the mixing-chamber stage, the device is enclosed in superconducting and cryopermalloy magnetic shields. Thermal noise is suppressed by cryogenic attenuators installed at multiple temperature stages. The flux-control lines are additionally equipped with homemade lossy powder filters, enabling both dc flux biasing and microwave control without introducing excess thermal photons.

The readout signal is first amplified by an impedance-matched Josephson parametric amplifier (IMPA), followed by a high-electron-mobility transistor (HEMT) amplifier at the 4-K stage and additional room-temperature amplifiers before digitization. The IMPA is operated using a dedicated microwave pump together with a separate precision dc current source for biasing. Cryogenic circulators and microwave filters suppress amplifier back-action and prevent thermal noise from propagating toward the device.

Control and readout waveforms are generated by 2.0~GSa/s arbitrary waveform generators. The fluxonium qubits are driven directly by the generators, while the coupler-control and readout signals are up-converted using homemade room-temperature IQ conversion modules driven by external microwave local oscillators. The IQ conversion module additionally provides the dc bias current required for coupler flux tuning. After amplification, the readout signal is down-converted by an IQ conversion module and digitized by 2.5~GSa/s analog-to-digital converters.

\section{Fidelity definition}

We use a common process-fidelity formalism to evaluate the effects of amplitude and phase calibration errors as well as qubit decoherence. Following the standard approach of Ref.~\cite{pedersen2007fidelity}, the implemented gate is described as a completely positive trace-preserving (CPTP) map

\begin{equation}
\mathcal{G}(\rho)=\sum_k G_k \rho G_k^\dagger,
\end{equation}
where $\{G_k\}$ are Kraus operators satisfying

\begin{equation}
\sum_k G_k^\dagger G_k=I.
\end{equation}

The average fidelity with respect to the ideal operation $U_0$ is

\begin{equation}
\label{AverFidelity}
F=
\frac{1}{d(d+1)}
\left[
\mathrm{Tr}
\left(
\sum_k M_k^\dagger M_k
\right)
+
\sum_k
\left|
\mathrm{Tr}(M_k)
\right|^2
\right],
\end{equation}
where $M_k=P U_0^\dagger G_k P^T$, $P$ is the projector onto the computational subspace, and $d$ is its dimension.

\section{Conditional phase calibration}

Here we derive the optimal calibration criterion for the conditional phase. For the implemented gate, the free parameters are the phase $\varphi_{111}$ accumulated by the computational state $\ket{111}$ and the virtual single-qubit $Z$ rotations, while the residual two-qubit phases are treated as coherent errors. The optimal calibration criterion is obtained by maximizing the average gate fidelity, Eq.~(\ref{AverFidelity}), with respect to these free parameters.

Let $\delta\varphi_i$ denote the deviation of the phase accumulated by computational state $i$ from its target value. Substituting the corresponding gate matrix into Eq.~(\ref{AverFidelity}) yields

\begin{equation}
F
=
\frac{1}{9}
+
\frac{1}{72}
\sum_{i,j}
\cos(\delta\varphi_i-\delta\varphi_j)
\approx
1-
\frac{1}{144}
\sum_{i,j}
(\delta\varphi_i-\delta\varphi_j)^2,
\label{eq:phase_F_global}
\end{equation}
where the summation is performed over all computational basis states.

Maximizing Eq.~(\ref{eq:phase_F_global}) with respect to the virtual single-qubit $Z$ rotations gives

\begin{equation}
\varphi_i
=
\frac{1}{4}
\left(
\sum_{j\notin\Omega_i}
\delta\varphi_j
-
\sum_{j\in\Omega_i}
\delta\varphi_j
\right),
\label{eq:opt_Z_rot}
\end{equation}
where $\Omega_i$ denotes the set of computational states with qubit F$_i$ in the excited state.

Substituting Eq.~(\ref{eq:opt_Z_rot}) into the remaining condition with respect to $\delta\varphi_{111}$ gives

\begin{equation}
2\delta\varphi_{111}
-
\delta\varphi_{110}
-
\delta\varphi_{101}
-
\delta\varphi_{011}
+
\delta\varphi_{000}
=
0,
\label{eq:opt_phi_111}
\end{equation}
which determines the phase that should be calibrated experimentally.

The required phase combination can be obtained from four Ramsey measurements:

(i) measuring qubit F$_1$ with qubits F$_2$ and F$_3$ prepared in $\ket{11}$, yielding
$\Delta_{\mathrm{i}}=\varphi_{111}-\varphi_{011}$;

(ii) measuring qubit F$_2$ with qubits F$_1$ and F$_3$ prepared in $\ket{11}$, yielding
$\Delta_{\mathrm{ii}}=\varphi_{111}-\varphi_{101}$;

(iii) measuring qubit F$_1$ with qubits F$_2$ and F$_3$ prepared in $\ket{10}$, yielding
$\Delta_{\mathrm{iii}}=\varphi_{110}-\varphi_{010}$;

(iv) measuring qubit F$_2$ with qubits F$_1$ and F$_3$ prepared in $\ket{00}$, yielding
$\Delta_{\mathrm{iv}}=\varphi_{010}-\varphi_{000}$.

Combining these measurements gives

\begin{equation}
\Delta_{\mathrm{i}}
+
\Delta_{\mathrm{ii}}
-
\Delta_{\mathrm{iii}}
-
\Delta_{\mathrm{iv}}
=
2\varphi,
\label{eq:calib}
\end{equation}

where $\varphi$ is the target conditional phase of the $\mathrm{CCPhase}$ gate. This choice of Ramsey measurements is not unique, and any equivalent set yielding the same linear combination of phases can be used. The derivation is valid for an arbitrary $\mathrm{CCPhase}(\varphi)$ gate, since only the phase deviations from their target values enter the analysis. Calibration of the $\overline{\mathrm{CCP}}\mathrm{hase}$ gate is performed analogously by inverting the initial computational states and reversing the sign of the measured phase differences.

\section{Sensitivity to calibration errors}
\subsection{Amplitude}

We first consider the effect of an imperfect drive-amplitude calibration. An amplitude error results in incomplete population return of the coupler after the gate pulse. The implemented gate differs from the ideal CCZ only in the diagonal element corresponding to the $\ket{111}$ state, which becomes $-1+\delta$. Substituting the resulting process matrix into Eq.~(\ref{AverFidelity}) yields

\begin{equation}
F=1-\frac{\delta^2}{4}.
\end{equation}

To achieve fidelities of 99.9\% and 99.99\%, the amplitude deviation must not exceed $\delta=0.06$ and $\delta=0.02$, respectively. For the 65-ns gate, whose calibrated pulse amplitude is approximately 0.35~V, these thresholds correspond to amplitude calibration deviations of about 26~mV and 13~mV.

\begin{figure*}
    \centering
    \includegraphics[width=\linewidth]{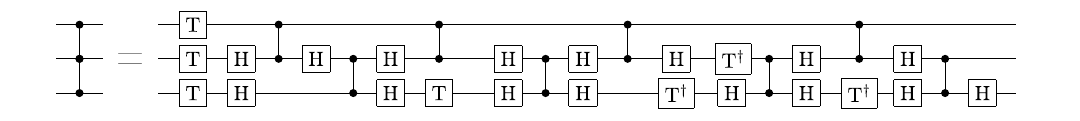}
    \caption{\textbf{CCZ decomposition into CZ gates for linearly connected qubits.}}
    \label{fig:CCZ_decomposition}
\end{figure*}

\subsection{Frequency}

We next consider the effect of an error in the accumulated conditional phase, such that the $\ket{111}$ state acquires a phase of $\pi+\delta\varphi$ instead of the ideal value $\pi$. Substituting the corresponding gate matrix into Eq.~(\ref{AverFidelity}) yields

\begin{equation}
    F = \frac{1}{9}+\frac{25+7\cos{\delta\varphi}}{36}\approx 1-\frac{7}{72}\delta\varphi^2.
\end{equation}

In the experiment, virtual $Z$ rotations are applied to the three fluxonium qubits. These rotations redistribute the accumulated phase error among the computational basis states. Applying Eq.~(\ref{eq:opt_Z_rot}) yields the optimal virtual Z rotations:

\begin{equation}
    \varphi_i = -\frac{\delta\varphi}{4}
\end{equation}

The corresponding fidelity then becomes

\begin{equation}
    F \approx 1-\frac{\delta\varphi^2}{18}.
\end{equation}

For the 65-ns gate, the corresponding phase-error thresholds translate into frequency calibration deviations of approximately 2~MHz and 600~kHz for fidelities of 99.9\% and 99.99\%, respectively.

\section{Coherence limit}

To estimate the coherence limit shown in Fig.~3a of the main text, we apply the same formalism (Eq.~\eqref{AverFidelity}) to energy relaxation and pure dephasing, obtaining

\begin{equation}
\label{3Q_fidelity}
F=
1-
\frac{4\tau}{9}
\sum_{i\in\{F_1,F_2,F_3\}}
\left(
\frac{1}{T_{1,i}}
+
\frac{1}{T_{\varphi,i}}
\right),
\end{equation}
where $\tau$ is the gate duration, while $T_{1,i}$ and $T_{2,i}^{\mathrm{echo}}$ are obtained from independent relaxation and spin-echo measurements (Table~I in the main text). The pure-dephasing time is calculated as

\begin{equation}
\frac{1}{T_{\varphi,i}}
=
\frac{1}{T_{2,i}^{\mathrm{echo}}}
-
\frac{1}{2T_{1,i}}.
\end{equation}

\begin{table}[t]
\centering
\caption{\textbf{Single-qubit operations.} 
Average fidelities of single-qubit operations measured individually and simultaneously by XEB. Each operation has a duration of 72~ns and is implemented using two 36-ns $\sqrt{X}$ gates combined with virtual $Z$ rotations.}
\label{tab:1Q_operations}
\begin{tabular}{lccc}
\hline
\hline
 & \textbf{F1} & \textbf{F2} & \textbf{F3} \\
\hline
$F_{\mathrm{ind}}$ $(\%)$ & 99.65 & 99.78  & 99.73  \\
$F_{\mathrm{sim}}$ $(\%)$ & 99.68 & 99.74  & 99.67  \\
\hline
\hline
\end{tabular}
\end{table}

\section{CCZ gate benchmarking with single-qubit reference}

The fidelity of the three-qubit gate is characterized using interleaved XEB with single-qubit reference sequences, a protocol widely used for the characterization of non-Clifford gates~\cite{Foxen_2020, chen2025efficient}.
As shown in Ref.~\cite{simakov2026decay}, the interleaved sequences with the CCZ gate from a circuit depth of four and above provide sufficient randomization and follows the standard multi-qubit depolarizing model. When local noise dominates, the reference decay parameter becomes  
\begin{equation}
p_{\mathrm{ref}} = 1-\frac{16}{21}(e_1+e_2+e_3),
\label{eq:gate_decay}
\end{equation}
where $e_i=1-p_i$ is the depolarizing error of the $i$th qubit. The depolarizing parameter of the target CCZ gate is then obtained using the conventional relation $p_{\mathrm{CCZ}}=p_{\mathrm{int}}/p_{\mathrm{ref}}$, yielding the corresponding gate fidelity $F_{\mathrm{CCZ}}=(7p_{\mathrm{CCZ}}+1)/8$. Here $p_{\mathrm{int}}$ is extracted from the interleaved XEB decay. The validity of the local-noise assumption is verified by comparing the simultaneous and individual single-qubit gate fidelities (Table~\ref{tab:1Q_operations}) and by the agreement between the measured reference decay and the analytical model on the same processor as demonstrated in Ref.~\cite{simakov2026decay}.

\section{Simulation of the CCZ decomposition}

To compare the demonstrated native $\mathrm{CCZ}$ gate with decomposition-based implementations, we simulate the standard Toffoli-equivalent decomposition for linearly connected qubits shown in Fig.~\ref{fig:CCZ_decomposition}, which consists of eight CZ gates~\cite{nielsen2010quantum}.
The experimentally realized $\overline{\mathrm{CCZ}}$ gate differs from the canonical $\mathrm{CCZ}$ only by local single-qubit basis transformations, these operations are omitted in the simulation.
The scheme is simulated using the superconducting-inspired SI1000 circuit-level noise model~\cite{Gidney_2021}. Each CZ gate is assigned a depolarizing error probability $p$, while Hadamard gates are assigned a depolarizing error probability $p/10$. The $T$ gates are treated as ideal, since they can be implemented virtually. During each two-qubit gate, qubits that remain idle are subjected to independent single-qubit depolarizing errors with probability $p/10$.
The CZ gate error probability $p$ is varied to determine the equivalent two-qubit gate fidelity required for the decomposition to reproduce the experimentally measured $\mathrm{CCZ}$ fidelity. The resulting dependence is shown in Fig.~3c of the main text.

\normalem{}
\bibliography{main}

\end{document}